\documentclass[sigconf,natbib=false]{acmart}

\makeatletter
\def\input@path{{ACM/}{./}}
\makeatother
\usepackage[table]{colortbl}
\usepackage{amsmath}
\usepackage{textcomp}
\usepackage{stfloats}
\usepackage{url}
\usepackage{graphicx}
\usepackage{multirow}
\usepackage{float}
\usepackage{eucal}
\usepackage[inline]{enumitem}
\usepackage{subcaption}
\usepackage[linesnumbered, ruled, vlined]{algorithm2e}
\usepackage{listings}
\usepackage{pifont}
\AtBeginDocument{%
  }

\acmYear{2024}\copyrightyear{2024}
\setcopyright{rightsretained}
\acmConference[MMSys '24]{ACM Multimedia Systems Conference 2024}{April 15--18, 2024}{Bari, Italy}
\acmBooktitle{ACM Multimedia Systems Conference 2024 (MMSys '24), April 15--18, 2024, Bari, Italy}
\acmDOI{10.1145/3625468.3652194}
\acmISBN{979-8-4007-0412-3/24/04}

\RequirePackage[
  datamodel=acmdatamodel,
  style=acmnumeric,
  sorting=none,
  ]{biblatex}

\begin{document}

\balance

\title{PyStream: Enhancing Video Streaming Evaluation}
\author{Samuel Radler}
\authornote{Both authors contributed equally to this research.}
\affiliation{
  \institution{Alpen-Adria-Universit{\"a}t Klagenfurt, Christian Doppler Laboratory ATHENA}
  \streetaddress{Universit{\"a}tstrasse 65/67}
  \city{Klagenfurt}
  \country{Austria}
}

\author{Leon Prüller}
\authornotemark[1]
\affiliation{
  \institution{Alpen-Adria-Universit{\"a}t Klagenfurt, Christian Doppler Laboratory ATHENA}
  \streetaddress{Universit{\"a}tstrasse 65/67}
  \city{Klagenfurt}
  \country{Austria}
}

\author{Emanuele Artioli}
\affiliation{
  \institution{Alpen-Adria-Universit{\"a}t Klagenfurt, Christian Doppler Laboratory ATHENA}
  \streetaddress{Universit{\"a}tstrasse 65/67}
  \city{Klagenfurt}
  \country{Austria}
}

\author{Farzad Tashtarian}
\affiliation{
  \institution{Alpen-Adria-Universit{\"a}t Klagenfurt, Christian Doppler Laboratory ATHENA}
  \streetaddress{Universit{\"a}tstrasse 65/67}
  \city{Klagenfurt}
  \country{Austria}
}

\author{Christian Timmerer}
\affiliation{
  \institution{Alpen-Adria-Universit{\"a}t Klagenfurt, Christian Doppler Laboratory ATHENA}
  \streetaddress{Universit{\"a}tstrasse 65/67}
  \city{Klagenfurt}
  \country{Austria}
}

\renewcommand{\shortauthors}{Radler, Prüller, et al.}

\newcommand{\etal}{\emph{et~al.}}
\newcommand{\ie}{\emph{i.e.}, }
\newcommand{\eg}{\emph{e.g.}, }
\newcommand{\etc}{\emph{etc.}}
\newcommand{\HAS}{\emph{HTTP Adaptive Streaming }}
\newcommand{\has}{HAS}
\newcommand{\DASH}{\emph{Dynamic Adaptive Streaming over HTTP }}
\newcommand{\HLS}{\emph{HTTP Live Streaming }}

\begin{abstract}

As streaming services become more commonplace, analyzing their behavior effectively under different network conditions is crucial. 
This is normally quite expensive, requiring multiple players with different bandwidth configurations to be emulated by a powerful local machine or a cloud environment. 
Furthermore, emulating a realistic network behavior or guaranteeing adherence to a real network trace is challenging.
This paper presents PyStream, a simple yet powerful way to emulate a video streaming network, allowing multiple simultaneous tests to run locally. By leveraging a network of Docker containers, many of the implementation challenges are abstracted away, keeping the resulting system easily manageable and upgradeable. 
We demonstrate how PyStream not only reduces the requirements for testing a video streaming system but also improves the accuracy of the emulations with respect to the current state-of-the-art. On average, PyStream reduces the error between the original network trace and the bandwidth emulated by video players by a factor of 2-3 compared to Wondershaper, a common network traffic shaper in many video streaming evaluation environments. Moreover, PyStream decreases the cost of running experiments compared to existing cloud-based video streaming evaluation environments such as CAdViSE.
\end{abstract}

\ccsdesc[500]{Networks~Network simulations}
\ccsdesc[300]{Networks~Network performance modeling}
\ccsdesc[100]{Networks~Network experimentation}
\ccsdesc[300]{Networks~Network performance analysis}

\keywords{Video Streaming Evaluation, HTTP Traffic Shaper, Evaluation Cost}

\maketitle

\section{Introduction} \label{Introduction}
Video streaming is by far the most data-intensive application of the Internet, accounting for approximately 65\% of the global network traffic~\cite{sandvine}. Therefore, it is crucial to develop tools for bandwidth management, as any 1\% increase in video streaming efficiency will save almost 1 billion TBs of Internet traffic~\cite{statista}. Adaptive video streaming is the current state-of-the-art multimedia delivery mechanism designed to optimize the viewer's experience by dynamically adjusting video quality based on real-time network conditions and device capabilities~\cite{tashtarian2022hxl3}. At its core, this technology leverages HTTP-based protocols and employs a segmented representation approach~\cite{tashtarianOscar}. Video content is encoded into discrete segments, each representing a fraction of the entire video stream. Leveraging adaptive bitrate algorithms (ABRs), video players dynamically select each segment's appropriate representation (bitrate and resolution), targeting crisp, seamless playback. By constantly monitoring network metrics such as available bandwidth and latency, ABRs autonomously switch between representations, striking a balance between video quality and rebuffering~\cite{nguyen2022dofp+}. 

\begin{figure*}
\centering
\begin{minipage}{.5\textwidth}
  \centering
  \includegraphics[width=0.9\linewidth]{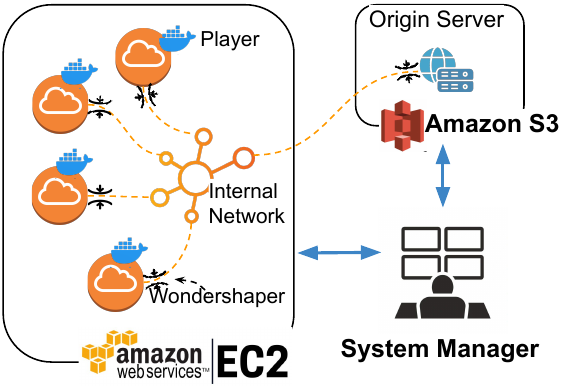}
  \captionof{figure}{CAdViSE architecture}
  \label{fig:CAdViSE}
\end{minipage}%
\begin{minipage}{.5\textwidth}
  \centering
  \includegraphics[width=0.8\linewidth]{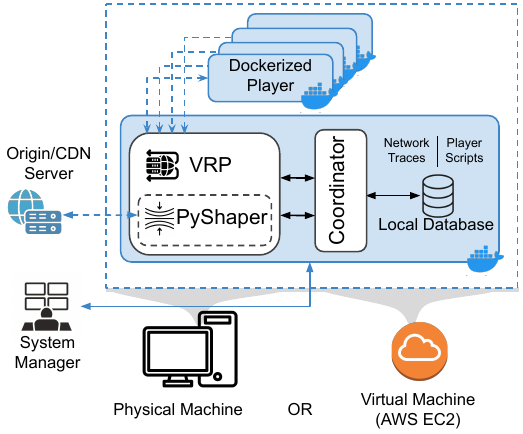}
  \captionof{figure}{Proposed PyStream architecture}
  \label{fig:PyStream}
\end{minipage}
\end{figure*}

Evaluating the performance of a video streaming pipeline (codec, ABR, etc.) requires monitoring many video playbacks in many combinations of network conditions, players, devices, bitrate ladder, etc.~\cite{tashtarian2024artemis}. This comes at a potentially prohibitive cost unless the testbed used is efficient in terms of cost and accuracy. Such a testbed must emulate multiple clients, each with a different player and a dedicated network trace approaching a real streaming scenario. In this way, monitoring streaming components' behavior can be parallelized, and performance can be evaluated efficiently. 
For example, CAdViSE~\cite{cadvise} is a cloud-based adaptive video streaming evaluation framework for the automated testing of adaptive media players (see Figure~\ref{fig:CAdViSE}). To run an experiment in CAdViSE with $n$ number of players, we need to instantiate $n$ EC2 machines and load a video player script (\eg dashjs~\cite{dash.js}, shaka~\cite{shaka}) in each. Therefore, the evaluation cost of CAdViSE is proportional to the number of EC2 instances used. Moreover, in this context, adhering to realistic network traces is crucial. For example, using an inefficient network traffic shaper may not generalize well for an ABR in the real world, leading to sub-optimal performance and low user satisfaction.

Wondershaper~\cite{wondershaper} is a network shaper used not only in CAdViSE but also in many other emulators that need to control network speed~\cite{8241127, 10.1145/2348688.2348703}. Wondershaper shapes the network bandwidth by connecting directly to a computer's network interface card (NIC) and bottlenecking it to the required bandwidth using the Linux built-in command, \ie the TC command~\cite{tc}.

In this paper, we address the cost and accuracy of video streaming emulators and introduce PyStream, a simple yet powerful emulator tool that initiates and monitors multiple players on a physical machine or a virtual machine (\eg AWS EC2~\cite{ec2}). PyStream is a docker-based emulator that, as Figure~\ref{fig:PyStream} shows, allows developers and researchers to run different numbers of players with various player types and network traces. 

Section \ref{Related Work} introduces the current state-of-the-art landscape regarding network shapers and video streaming monitoring tools and emulators. Section \ref{Problem Description and Motivation} describes the current problems with network emulation and motivates our work. Section \ref{PyStream Details} delves into the specific design and implementation of PyStream, while Section \ref{Performance} provides benchmarks and observations about PyStream's performance. Finally, Section \ref{Conclusion} concludes our work with use cases and conclusions.

\section{Related Work} \label{Related Work}

Regarding the defined objectives for designing PyStream, we have identified several tools that handle one or more steps in the traffic emulation pipeline.\par

\text{Wondershaper}~\cite{wondershaper} is a foundational network shaping tool that was released in 2002 and is still relevant nowadays. Conceptually simple, it is a script that allows the user to limit the bandwidth of one or more network adapters. It uses Linux's iproute TC command~\cite{tc} but greatly simplifies its operation.
\text{Mahimahi}~\cite{mahimahi} is another network emulation and traffic shaping framework that works under the record-replay principle. HTTP traffic can be recorded, stored, and replayed with different network conditions. For this matter, Mahimahi provides several tools, such as the RecordShell and ReplayShell for recording and replaying traffic, respectively, the DelayShell for emulating delays and the LinkShell to emulate packet loss. 
Trickle~\cite{eriksen2005trickle} is a tool for rate limiting TCP connections in smaller, unmanaged networks that runs in user-space. It utilizes the preloading functionality of the Unix dynamic loader to load its own socket library wrappers. It enables Trickle to delay and truncate socket I/Os without requiring administrator privileges; and, in this way it limits the bandwidth.
Dummynet~\cite{dummynet} is a network emulation tool widely used in the field of networking research. It includes a traffic shaper and a packet scheduler, enables the emulation of a whole set of network environments, and it has the ability to model delay, and packet loss, it is easy to integrate and provides consistent results.
With Mininet~\cite{mininet}, users can create, configure, and connect virtual network nodes using lightweight containerization technologies. This enables the emulation of complex network topologies on a single computer, in accordance with Mininet's focus on SDN research, education, and testing.
ns-3~\cite{ns-3} is a discrete-event network simulator targeted at research and education, allowing the modeling and simulation of computer networks with a focus on network protocols and scenarios.
\text{Sabre}~\cite{sabre} is an open-source framework that emulates real-world conditions to evaluate ABRs via quality of experience (QoE) metrics. Sabre can also be used to test new ABR algorithms without learning the intricacies of implementing the production layer. \text{AdViSE}~\cite{advise} is a framework that enables testing media players under various network conditions. It accomplishes this by creating software-defined networks that are then shaped into the desired network. This network is then hosted on a physical machine so the players can access it. \text{CAdViSE}~\cite{cadvise} improves over AdViSE by removing the system's physical deployment. This testbed can be operated on cloud-based systems such as Amazon AWS, making upscaling easier and standardizing the machine's performance. This architecture is continuously updated with new features and improvements. One such improvement is LLL-CAdViSE\cite{lllc}, which enables the testing of low-latency metrics. \par

Table~\ref{table:RelatedWorkOverview} presents a quick overview of video streaming emulators and network traffic shaper tools.

\begin{table}[h] 
\caption{Overview of related work over different metrics. (TS: traffic shaper, VSE: video streaming emulator, NE: network emulator, SE: switch emulator, and NS: network simulator)}
\begin{tabular}{ lcccc } 
  \hline
   & Tool & TS Type &  Platform\\ 
  \hline
  \hline

  Wondershaper~\cite{wondershaper} & TS & - &  Linux\\ 
  \hline
  Trickle~\cite{eriksen2005trickle} & TS & - &  Unix-like\\ 
  \hline
  Mahimahi~\cite{mahimahi} &NE, TS & - &  Linux/Windows\\ 
  \hline
  Dummynet~\cite{dummynet} & NE, TS & - &  Unix-like\\ 
  \hline
  Mininet~\cite{mininet} & SE & iproute TC &  Linux\\ 
  \hline
  ns-3~\cite{ns-3} & NS & iproute TC &  Unix-like\\ 
  \hline
  Sabre~\cite{sabre} &VSE & Wondershaper & Linux\\ 
  \hline
  AdViSE~\cite{advise} &VSE & Wondershaper & Linux\\ 
  \hline
  CAdViSE~\cite{cadvise} & VSE & Wondershaper &  Linux\\ 
  \hline
  \textbf{PyStream}& VSE & PyShaper &  Linux/Windows\\
 \hline
\end{tabular}
\label{table:RelatedWorkOverview}
\end{table}

\section{Problem Description and Motivation} \label{Problem Description and Motivation}
As mentioned earlier, PyStream aims to tackle two main challenges in video streaming emulators: the cost of emulation/evaluation and the need for accuracy. In Figure~\ref{fig:CAdViSE}, we illustrated that as the number of players increases in CAdViSE, the cost also escalates. This increase occurs because we have to assign an EC2 virtual machine to each player, pulling its image from the Docker hub. The limitation arises from Wondershaper~\cite{wondershaper}, the traffic shaper used in CAdViSE, which can only run on a physical or virtual network interface. Note that Wondershaper has been used as the main traffic shaper module in many emulators (see Table~\ref{table:RelatedWorkOverview}).\par
\begin{figure*}[t]
\includegraphics[width=\linewidth]{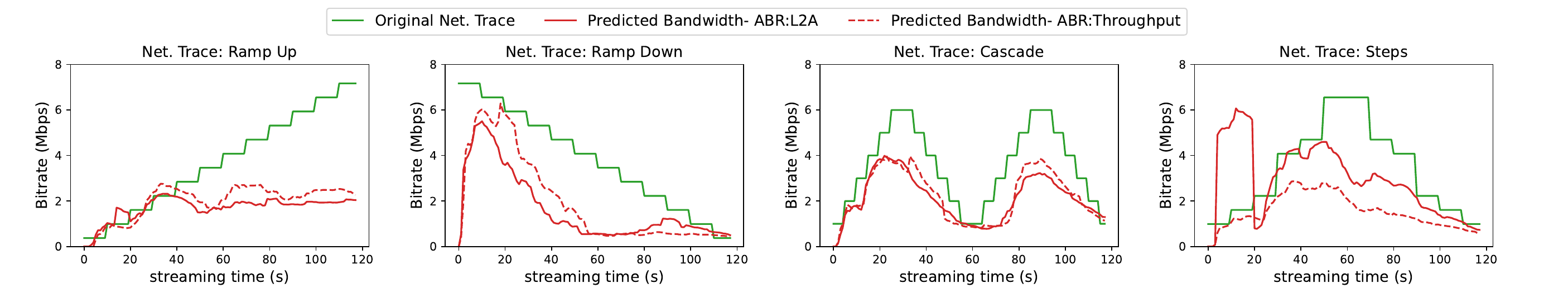}
\centering
\caption{Comparing the original network traces (Ramp Up, Ramp Down, Cascade, and Steps)~\cite{cadvise} with the predicted bandwidth by different ABRs: L2A~\cite{l2a} and throughput based~\cite{bentaleb2022low}.}
\label{motive}
\end{figure*}

From an accuracy standpoint, we compare the original network trace with the predicted bandwidth across various network traces\footnote{The Ramp Up network trace cycles through the values \{372, 990, 1608, 2226, 2844, 3462, 4080, 4698, 5316, 5934, 6552, 7170\} (kbps) after every 10 seconds, with 80 ms of constant latency. The Ramp Down network trace cycles through the same values as Ramp Up, in inverse order after every 10 seconds, with 80 ms of constant latency. The Cascade network trace cycles through the values \{1, 2, 3, 4, 5, 6, 5, 4, 3, 2, 1\} (Mbps) after every 5 seconds, with 80 ms of constant latency. The Steps network trace cycles through the values \{990, 1608, 2226, 4080, 4698, 6552, 4698, 4080, 2226, 1608, 990\} (kbps), after every 10 seconds, with 80 ms of constant latency.}~\cite{cadvise} and two ABR algorithms: L2A~\cite{l2a} and throughput based~\cite{bentaleb2022low}. 
As depicted in Figure~\ref{motive}, the predicted bandwidth by both ABR algorithms is not accurate. This behavior could be attributed to either the inaccuracy of the ABR bandwidth prediction module or the low performance of Wondershaper. While numerous studies have focused on optimizing the bandwidth prediction module in ABRs~\cite{bentaleb2022bob}, the primary reason for this behavior is the low accuracy in shaping network traffic with Wondershaper. 
The evaluation section will demonstrate that ABRs can accurately predict the bandwidth if paired with an efficient traffic shaper. Therefore, in this paper, we introduce PyStream, a Docker-based video streaming emulator which includes an efficient Python-based network traffic shaper called PyShaper. Our system is developed not only to enhance the accuracy of traffic shaping but also to reduce the cost of streaming emulation.

\section{PyStream Details} \label{PyStream Details}
Figure~\ref{fig:PyStream} illustrates the conceptual architecture of PyStream, which can be executed on either a physical or virtual (\eg AWS EC2) machine. PyStream operates through Docker and consists of the following modules: \textit{coordinator}, \textit{virtual reverse proxy} (VRP), \textit{PyShaper}, and the \textit{local database}. Before describing the details of these modules, let us explain how PyStream can be configured to run various video streaming scenarios. 

The system manager, orchestrating the experiments, first defines the quartet $\texttt{<[PID], PT, NT, L>}$, where \texttt{[PID]} is the list of player IDs that should run the player type \texttt{PT} (\eg dashjs~\cite{dash.js}, shaka~\cite{shaka}, etc.) operating over the network trace \texttt{NT}, that is similar to the proposed JSON format in CAdViSE. Moreover, $L$ is the duration of the emulation in seconds. 
We note here that \texttt{PID} can be a unique name for each player. When a player sends an HTTP request, its PID is embedded in the URL. Thus, we can easily customize the manifest address for each player by adding the \texttt{PID}.
The quartets are defined and stored in the local database by the system manager. The main responsibility of the coordinator module is to fetch a docker from Docker Hub, update the determined player script ($PT$), and run the docker. The coordinator also instructs PyShaper to apply the assigned network traces ($NT$) to the players' traffic. Dockerized players are configured to connect to the VRP module to start streaming. The VRP plays as a man-in-the-middle between the players and the origin/CDN server. In the VRP, PyShaper adjusts the bandwidth for each player by introducing synthetic delay when sending HTTP packets, according to the determined network traces.

\begin{figure}[b]
\includegraphics[width=\columnwidth]{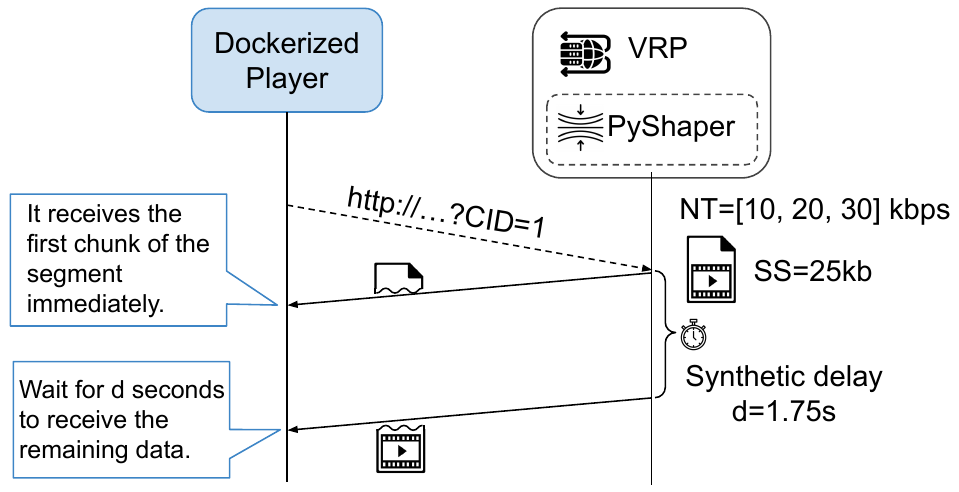}
\centering
\caption{An example of requesting a segment by a dockerized player in PyStream.}
\label{pyshaper}
\end{figure}

Whenever a player sends an HTTP request, the VRP downloads the requested segment from the origin/CDN server\footnote{In this study, we assume that all requested segments are available in the local database.}. The VRP then calculates the synthetic time the segment should have taken to download, according to the player's network trace and waits that long before forwarding the response to the player. To accurately follow the given network traces, a separate CPU thread updates the variables used for bandwidth calculations for each player every second. 
Crucially, before setting the thread to sleep for the calculated delay, part of the segment must be sent to ensure that the TCP connection remains active and an accurate bandwidth estimation at the player is maintained. Once the thread's sleep period concludes, the remaining segment content is transmitted to the player (see Figure~\ref{pyshaper}). 

Algorithm~\ref{algo_sleeptime} shows the logic for calculating the synthetic delay. To that end, it uses the following parameters:

\textbf{Input.}
\begin{itemize}
   \item Player ID (\texttt{PID}): This uniquely identifies the player that the algorithm is calling.
   \item Segment Size (\texttt{SS}): Represents the size of the currently requested segment by the player \texttt{PID}.
   \item Network Trace (\texttt{NT}): The set of network trace values used for the player \texttt{PID}.
   \item Streaming time (\texttt{ST}): A pointer that tracks the current position inside the set \texttt{NT} for the player \texttt{PID}.
\end{itemize}

\textbf{Global Variables.} Leftover Bandwidth (\texttt{LeftOverBW}) [global variables]: A two-dimensional array that, with the length of the number of players, stores unused bandwidth from their previously requested segments.

\textbf{Output.} Synthetic Delay (\texttt{d})[output]: The determined delay by the algorithm for player \texttt{PID}.

\begin{algorithm}[b]
\SetAlgoNlRelativeSize{0}
\SetKwInOut{Global}{Global}
\SetKwInOut{Input}{Input}
\SetKwInOut{Output}{Output}
\Global{LeftOverBW}
\Input{$PID$ (player ID), $SS$ (segment size), $NT$ (network trace), $ST$ (streaming time)}
\Output{$d$ (synthetic delay)}

$RBW \leftarrow SS\,/\,1\,\mathrm{sec}$ 

$ABW \leftarrow \mathrm{LeftOverBW}[PID]$

\eIf{$ABW \geq RBW$}{
    $ABW \leftarrow ABW - RBW$\\
    $\mathrm{LeftOverBW}[PID] \leftarrow ABW$\\
    $d \leftarrow SS\,/\,NT[ST]$\\
}{
    $d \leftarrow ABW\,/\,NT[ST]$\\
    $SS \leftarrow SS - ABW\times 1\,\mathrm{sec}$\\
    $ABW \leftarrow 0$

    \While{$RBW > ABW$}{
        $ST \leftarrow ST + 1$\\
        $ABW \leftarrow NT[ST]$\\
        $RBW \leftarrow SS\,/\,1\,\mathrm{sec}$

        \eIf{$ABW \geq RBW$}{
            $d \leftarrow d + RBW\,/\,ABW$\\
            $ABW \leftarrow ABW - RBW$\\
            $RBW \leftarrow 0$\\
            $\mathrm{LeftOverBW}[PID] \leftarrow ABW$\\
        }{
            $SS \leftarrow SS - ABW\times 1\,\mathrm{sec}$\\
            $d \leftarrow d + 1$
        }
    }
}
    \KwRet{$d$}
\caption{Synthetic Delay Calculator Function}
\label{algo_sleeptime}
\end{algorithm}

Let us demonstrate Algorithm~\ref{algo_sleeptime} through a simple example. Consider a dockerized player with \texttt{PID}=1 requesting a segment with a size of \texttt{SS}=25~kb (see Figure~\ref{pyshaper}). Additionally, we have a simple network trace \texttt{NT} which consists of three values \texttt{NT}=\{10, 20, 30\}~kbps, representing the available bandwidth over three seconds of streaming. It's important to note that if the values of \texttt{NT} are set for a specific duration longer than a second, we should scale \texttt{NT} down to a one-second interval by repeating its values according to its duration. We also set \texttt{LeftOverBW[1]}=0.
To calculate the correct delay for the requested segment with size \texttt{SS}, the algorithm first calculates the required bandwidth (\texttt{RBW}), which is 40~kb/1~sec (line 1). If bandwidth from an earlier call is available, it is stored in an available bandwidth variable, denoted by \texttt{ABW}. In the first call, we have \texttt{ABW = 0}. If enough bandwidth from the last call is available (the 'if' condition in line 3), then the synthetic delay \texttt{d} is calculated in line 6. However, if \texttt{ABW < RBW}, the 'else' path is executed.
In the first three lines of the 'else' block,  we initialize delay \texttt{d} using the remaining \texttt{ABW} from the last function call, reduce \texttt{SS} by \texttt{ABW} $\times$ 1~sec, and set \texttt{RBW}=0, in lines 8-10, respectively. 
The 'while' loop in line 11 is entered because \texttt{RBW > ABW}. \texttt{ST} is increased by one since we need to look for the next entry of the \texttt{NT} array and increase \texttt{ABW} accordingly. Thus, we have \texttt{ABW} = \texttt{NT[ST]} = 10~kbps.
With the current state of the values, the 'if' clause in line 15 is false, and the else block must be executed. Therefore, \texttt{SS} is decreased by \texttt{ABW} = 10 $\times$ 1~sec, which means \texttt{SS} is now 15, and \texttt{d} is increased by 1.

The next execution of the while loop in Line 11 will once again increase \texttt{ST} by 1, set \texttt{ABW} to the next network trace entry \texttt{NT[ST]}, and then recalculate \texttt{RBW}.
Now, the 'if' clause in line 15 is true, and then \texttt{d} is set to \texttt{d} = \texttt{d} + \texttt{RBW / ABW} = 1.75. It means that the segment transmission must be delayed by 1.75 seconds to emulate the selected network trace. The only thing left is to reduce the \texttt{ABW} by the \texttt{RBW} that was just used. This value is then saved into \texttt{LeftOverBW[1]} followed by setting \texttt{RBW} to 0. Since \texttt{RBW} is now 0, the while loop terminates in the next iteration, and \texttt{d} is returned in line 23.

\section{Performance} \label{Performance}

In order to assess PyStream's performance, we setup an AWS EC2 instance of type c6a.32xlarge, which is equipped with the following components:
\begin{itemize}
    \item CPU: 64 cores @ 3.6 GHz with hyperthreading, for a total of 128 threads
    \item RAM: 256 GB
    \item OS: Ubuntu 20.04 LTS
\end{itemize}

The machine's performance is monitored with the built-in Docker Stats~\cite{docker-stats}, and a custom script logged each player's load at every second. To benchmark these results, the same tests are run on CAdViSE, in order to emulate the same traces used in PyStream.

\subsection{Accuracy Tests}

\begin{figure*}[t]
\includegraphics[width=\linewidth]{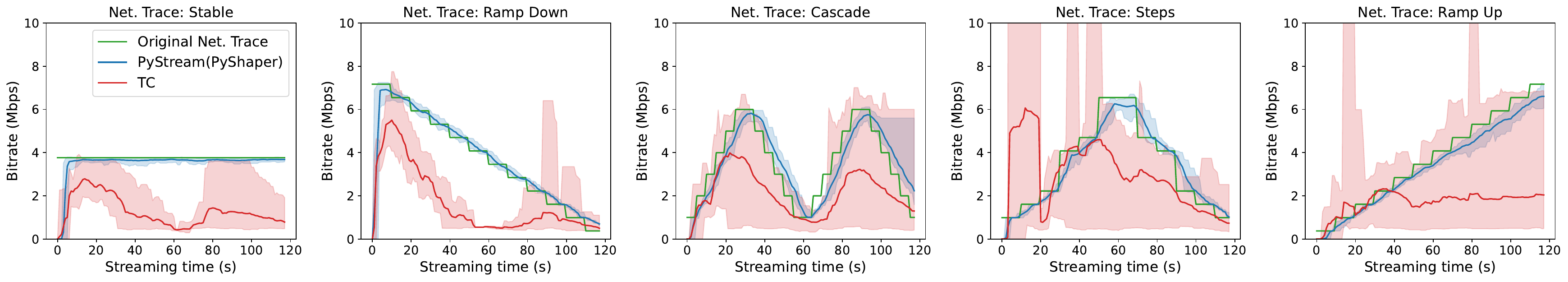}
\centering
\caption{Comparing the variation between the original network trace and predicted bandwidth by L2A ABR using PyStream and Wondershaper to shape the network traffic across various network traces.}
\label{acc-l2a}
\end{figure*}

\begin{figure*}[t]
\includegraphics[width=\linewidth]{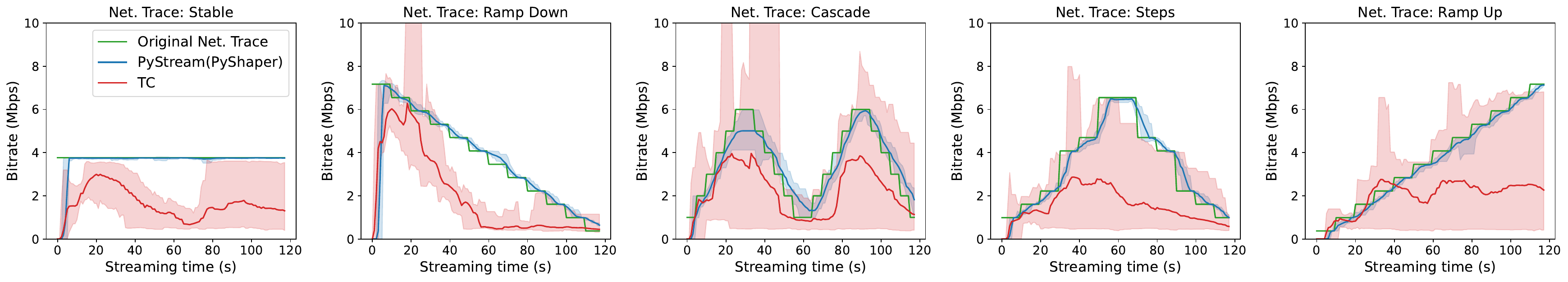}
\centering
\caption{Comparing the variation between the original network trace and predicted bandwidth by Throughput ABR using PyStream and Wondershaper to shape the network traffic across various network traces.}
\label{acc-throughput}
\end{figure*}
\begin{figure*}[t]
\includegraphics[width=\linewidth]{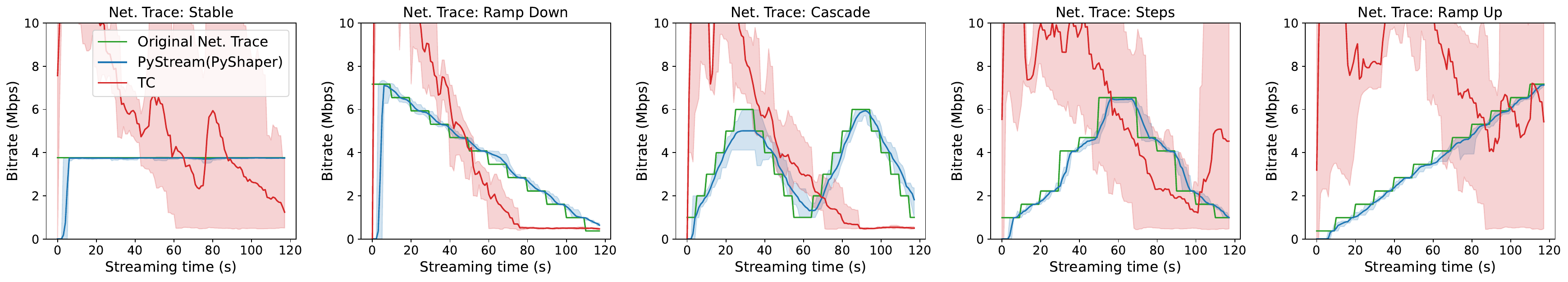}
\centering
\caption{Comparing the variation between the original network trace and predicted bandwidth by Throughput ABR using PyStream and Linux TC command to shape the network traffic across various network traces. }
\label{acc-throughput-tc}
\end{figure*}
The accuracy tests are performed in order to see how close the players' predictions are to the real network traces. For this, 20 players run for 120 seconds, and the average bandwidth predictions of L2a and Throughput-based ABRs are plotted in Figures~\ref{acc-l2a} and~\ref{acc-throughput}, respectively. Around the average value, the range between minimum and maximum player value is shown as an indication of variability. It highlights how predicated bandwidth by ABRs is closer and much more consistent among each other to the original network trace when we use Pyshaper compared to the experiments using wondershaper.

To have a better evaluation on the performance of PyStream, in the next scenario, we use Linux TC command~\cite{tc} to shape the bandwidth for players. As shown in Figure~\ref{acc-throughput-tc}, PyStream efficiently emulates the bandwidth of the given network trace in comparison with TC tool. 
\subsection{Load Tests}
\begin{figure*}[t]
\includegraphics[width=\linewidth]{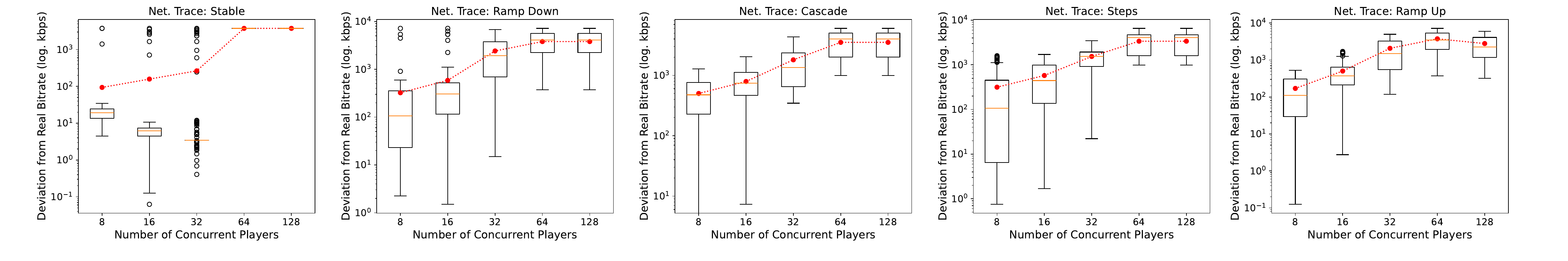}
\centering
\caption{MAE and Variance in relation to concurrent players}
\label{load plots}
\end{figure*}

Having shown the PyStream's accuracy on different ABRs and network traces, PyStream is then tested on its scalability. Specifically, how increasing the number of players would impact the accuracy. To this end, the experiment is repeated with 8, 16, 32, 64, and 128 players. The number of players was chosen due to the EC2 machine's resources to check the response of PyStream when the number of threads needed for players became equal (64) and higher (128) than the physical number of cores (64). Figure~\ref{load plots} shows the results of this stress test, where the red dot corresponds to the mean absolute error (MAE), calculated as follows:
\[MAE = \frac{\sum_{i=0}^{n}|T_{i}-t_{i}|}{n}\], where $T_{i}$ is the real bitrate for client i, $t_{i}$ is the bitrate predicted by the player for client i, and $n$ is the number of players.
The results confirm that, while the number of players is much lower than the number of physical cores, PyStream-caused deviations from the network traces remain small and increase exponentially (linearly in our log plot) up to the physical number of cores. Once the number of processes surpasses the number of cores, it appears that the performance stabilizes, but this is due to the machine being forced to freeze several threads. Therefore, performances above this threshold should not be considered.

\subsection{Resouce Tests}

\begin{figure}[t]
\includegraphics[width=\linewidth]{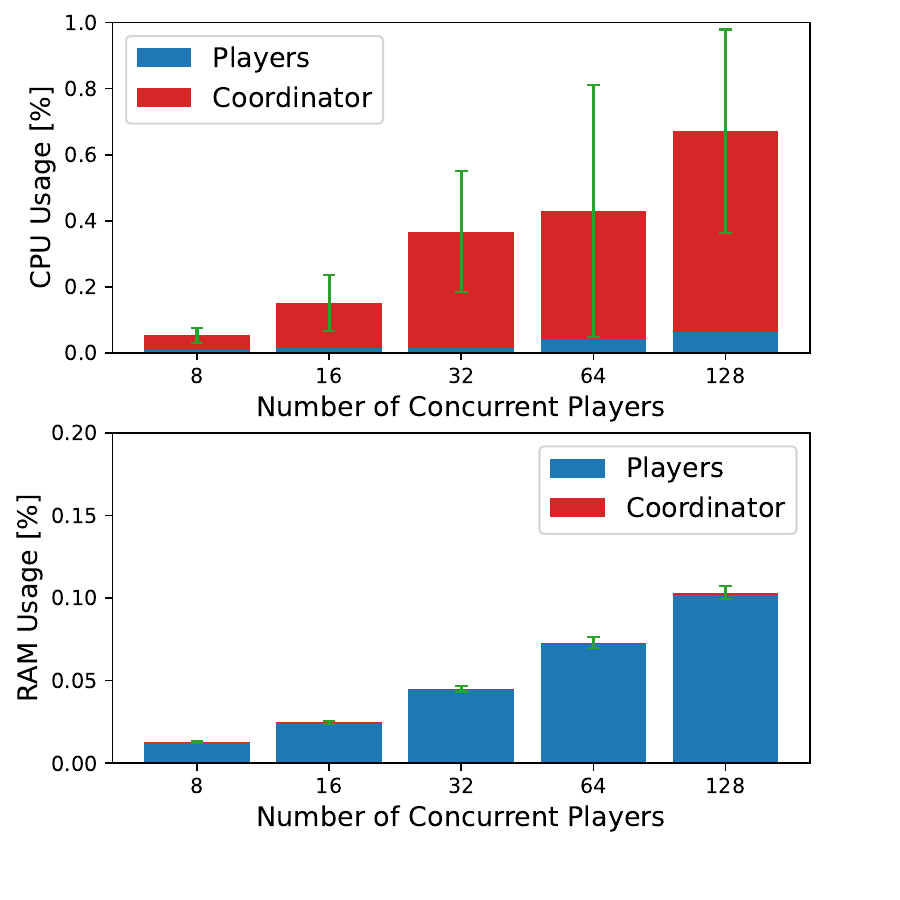}
\centering
\caption{Resource usage in relation to concurrent players}
\label{resource plot}
\end{figure}
In the final test, we investigate the connection between the machine's resources and PyStream's utilization of said resources. This is extrapolated from running the same experiments as above and logging each Docker container's ($n$ players + 1 controller) resource utilization over time.
Figure~\ref{resource plot} shows a linear dependence of both CPU and RAM usage to the number of players, but not from the same containers: while CPU usage is negligible for player containers but high for the controller, the opposite is true for RAM. This is because the controller is fundamentally a Python script that needs to run calculations for each player, while each player does not perform complex calculations but has to store a player and, crucially, its streamed video segments. 
This is a fundamental point in favor of PyStream's architecture over CAdViSE: typical virtual machines' CPU count and RAM scale together. Therefore, a system that is able to leverage both from the same machine is a more efficient use of resources than needing to spin up multiple machines of which only one dimension is fully utilized. 
As can be seen from Figure~\ref{resource plot}, the CPU count is the bottleneck of this system, while the required RAM is much lower than the physical limit of an EC2 machine with a sufficient number of cores. Therefore, both PyStream and CAdViSE need the same-sized machine to perform an experiment with the same number of players.

Quantifying the resource-saving is then a simple matter of summing the costs of EC2 machines required for CAdViSE's players. Figure~\ref{resource plot} shows how, for 64 players, the total utilization of RAM is around 7.5\% of the total 256 GB available, resulting in 0.3 GB of RAM per player. This amount of RAM is serviced already by the smallest EC2 instance type, namely t2.nano, which costs 0.6 cents/hour, for a total cost of 38 cents/hour. Given that a 64-core EC2 instance type such as costs about 5 euros/hour, the savings in this experiment amount to around 8\%. However, this calculation assumes that the t2.nano instance type is actually able to run a CAdViSE player instance and that CAdViSE's server requires similar computational power than PyStream's controller. Therefore, more experiments are necessary to quantify the cost comparisons between the two systems, and possibly others, with precision.

\section{Conclusion} \label{Conclusion}

In this study, we presented PyStream, a novel way to emulate network traces through a Docker environment, making it possible to run multiple simultaneous tests locally. We showed the limitations of current approaches, especially with regard to traffic shaping, and how PyStream significantly improves the accuracy of the emulations by developing PyShaper, an algorithm that calculates how long each segment should take to reach each player given their assigned network trace, and keeps the player from receiving it beforehand. Furthermore, by combining the resource requirements of the different parts of a network emulator, PyStream is able to make full use of a single virtual/physical machine, reducing the cost of running experiments. While further exploration and more comprehensive tests could provide improvements, we showed how PyStream's approach is a valuable addition to the current landscape of network emulation.

\printbibliography

\end{document}